\documentclass[final,5p,times,twocolumn,12pts]{elsarticle}
\usepackage[T1]{fontenc}
\usepackage{textcomp}
\usepackage{amsmath}
\usepackage{amssymb}
\usepackage{graphicx}
\usepackage[francais]{babel}
\usepackage[utf8]{inputenc}
\usepackage[T1]{fontenc}
\usepackage{natbib,hyperref}
\usepackage{wasysym}
\makeatletter

\providecommand{\tabularnewline}{\\}

\makeatother

\usepackage{babel}
\begin{document}

\journal{NIM-A}

\title{Modelling of laser-plasma acceleration of relativistic electrons in the frame of ESCULAP project
}

\ead{kubytsky@lal.in2p3.fr}

\author[CLUPS]{E.Baynard }

\author[LAL]{C. Bruni}
\author[LAL]{K. Cassou}
\author[LAL]{V. Chaumat}
\author[LAL]{N. Delerue}

\author[LPGP]{J.Demailly}

\author[LAL]{D.Douillet}
\author[LAL]{N. El Kamchi}

\author[CEA]{D. Garzella}
\author[CLUPS]{O. Guilbaud}

\author[LAL]{S. Jenzer}
\author[LPGP]{S. Kazamias}

\author[LAL]{V. Kubytskyi\corref{cor1}}
\author[LAL]{P. Lepercq}

\author[LPGP]{B. Lucas}
\author[LPGP]{G. Maynard}
\author[LPGP]{O. Neveu}

\author[CLUPS]{M. Pittman}
\author[CLIO]{R. Prazeres}

\author[LAL]{H. Purwar}
\author[CLUPS]{D. Ros}
\author[LAL]{K. Wang}


\address[LAL]{Laboratoire de l'Acc\'el\'erateur Lin\'eaire (LAL), Univ. Paris-Sud, CNRS/IN2P3, Universit\'e Paris-Saclay, Orsay, France}

\address[CLUPS]{Centre Laser de l'Universit\'e Paris-Sud (CLUPS), Univ. Paris-Sud, Universit\'e Paris-Saclay, Orsay, France}

\cortext[cor1]{Corresponding author}

\address[LPGP]{Laboratoire de Physique des Gaz et des Plasmas (LPGP), CNRS, Univ. Paris-Sud, Universit\'e Paris-Saclay, Orsay, France}

\address[CEA]{Laboratoire Interactions, Dynamiques et Lasers (LIDYL), CEA/DRF, Universit\'e Paris-Saclay, Saclay, France}
\address[CLIO]{Centre Laser Infrarouge d'Orsay, Laboratoire de Chimie Physique (CLIO/LCP), Univ. Paris-Sud, CNRS, Universit\'e Paris-Saclay, Orsay, France}

\date{\today}
\begin{abstract}
We present numerical simulations results on the injection and acceleration
of a 10 MeV, 10 pC electrons beam in a plasma wave generated in a
gas cell by a 2J, 45 fs laser beam. This modeling is related to the
ESCULAP  project in which the electrons accelerated by the PHIL photo-injector
is injected in a gas cell irradiated by the laser beam of the LASERIX
system. Extensive modeling of the experiment was performed in order
to determine optimal parameters of the laser plasma configurations.
This was done with the newly developed numerical code WakeTraj . We
propose a configuration that benefits of a highly compressed electron
bunch and for which the injected electron beam can be efficiently
coupled to the plasma wave and accelerated up to 140 MeV, with an
energy spread lower than 5\%.
\end{abstract}

\begin{keyword}
LPA  \sep Modeling  \sep WAKE-EP \sep WakeTraj
\end{keyword}

\maketitle

\section{\label{sec:level1}Introduction}

The objective of the ESCULAP (ElectronS CoUrts pour L'Acc\'el\'eration Plasma) project is the experimental study of
Laser-Plasma Acceleration (LPA) of an injected relativistic electron
bunch generated by a photo-injector. The experimental configuration
of ESCULAP is presented in \cite{NiDelerueEAAC2017} and the optimization
of the ESCULAP beam line is described in details in \cite{KeWangEAAC2017}.
Here we analyze, through numerical simulation, the coupling of the
electron beam, as produced by the ESCULAP beam line, with a plasma
wave generated in a gas cell by irradiation of the 2J laser beam of
the LASERIX system. A parametric study over many parameters has to
be performed in order to define an optimized configuration. Here we
present a first phase of this optimization process, where some simplifications
are introduced, such as a perfect gaussian laser beam and a uniform
density target.

The parametric study, which concerns a large number of parameters,
can require very heavy calculations. We have developed numerical tools
in order to perform more efficiently such parametric studies. This
will be discussed in the next section. In section 3 we will present
selected results concerning the influence of the laser focal plane
position and the delay between the laser and the electron beam from
which we proposed an optimized configuration.

\section{Numerical modeling}
\label{sec:numerical}
\subsection{Numerical setup}
Numerical modelings have being developed to optimize the experimental
configuration of the accelerating plasma cell in the framework of
the ESCULAP project through parametric studies. This modeling effort
is integrated within the simulation part of the ESCULAP project \cite{NiDelerueEAAC2017}
in order to get a global start-to-end numerical simulation of the
electrons beam dynamics. Considering the interaction of an injected
electron beam with a plasma wakefield, the properties of the injected
electrons beam are extracted from an ASTRA file related to the optimized
configurations of the longitudinally compressed bunch presented in
\cite{KeWangEAAC2017}. In these configurations, the electron bunch
has an average energy of 10 MeV, a total charge of 10 pC, a duration
(both for the RMS and FWHM values) close to 100 fs, and a transverse
size between 30 and 50 \textmu m at focus. Due to the relatively high
value of this transverse size, preliminaries studies have shown that
the electron bunch should be focused at the entrance of the plasma
cell. The laser properties are those of the LASERIX system (see
\cite{NiDelerueEAAC2017}), with a wavelength of $\lambda$ = 0.8
\textmu m, a maximum of energy of 2J and a FWHM duration of 45 fs,
leading to a maximum power of $P_{L}$= 41 TW, the laser profile being
assumed to have a Gaussian form both in its longitudinal and transverse
dimensions. The focalisation of laser is determined so as to have
a large enough waist, in order to capture the maximum of the injected
electrons but also to provide the highest intensity to get a maximum
of the accelerating field. Preliminary studies have shown that a waist
around 50 \textmu m at focus provides a good compromise. In the presented
results we fix the value of the waist at $w_{0}=$50.46 \textmu m
which corresponds to a Rayleigh length $Z_{R}=\pi w_{0}^{2}/\lambda$
of 1 cm and a maximum intensity in vacuum of $I_{0}=10^{18}$ cm$^{2}$.
The position of the laser focus plane is an important parameter, which
influence will be analyzed in the next section. A second important
parameter is the density of the plasma cell. Here, due to the high
intensity of the laser, we can assume that the plasma is fully ionized,
and to simplify the analysis, we assume a uniform density profile.
The equilibrium density of the plasma $n_{e0}$ is chosen so that
the plasma period $t_{p}$ is much larger than the duration of the
electron bunch, still keeping large transverse and longitudinal
fields in order to efficiently trap and accelerate the electrons. 
For all numerical studies in this paper we choose a density $n_{e0}=2\times10^{17}$ cm$^{-3}$, which offers a good compromise. At this density, $t_{p}$ = 249 fs, while the maximum
accelerating field, as predicted by the linear theory \cite{esarey_physics_2009},
is of 56 MeV/cm. At this density, the critical power for relativistic
self-focusing is three times larger than $P_{L}$, therefore relativistic
self-focusing does not play an important role. In fact the perturbation
parameter of $\delta n/n_{e0}$, with $\delta n$ the perturbation
of density induced by the interaction with the laser, has a maximum
value of 0.13, corresponding to the so-called quasi-linear interaction
domain, in which non-linear effects are expected to be small, but
non-negligible. This point will be addressed below.

The most accurate simulation tools related to LPA are based on the
Particle-In-Cell (PIC) method. 3D effects have to be taken into account
in LPA modeling and even with the most advances technics \cite{vay_recent_2016},
3D PIC simulations lead to very heavy calculations, preventing to
perform parametric studies over a large ensemble of parameters. In
our case however, some suitable approximations can be introduced in
order to reduce the computing time. The considered densities correspond
to under-dense plasmas, at which the three parameters $\varepsilon_{1}=\lambda_{p}/Z_{r}$,
$\varepsilon_{2}=\lambda/w_{0}$ and $\varepsilon_{3}=n_{e0}/n_{c}$, with
$n_{c}$ the critical density, are all much smaller than one : $\varepsilon_{1}=0.1\%$,
$\varepsilon_{2}=1.6\%$ and $\varepsilon_{3}=1.1\%$. For these conditions,
the physical model used in the WAKE numerical code \cite{mora_kinetic_1997},
where a quasi-static approach is used to described the plasma electrons
dynamics and an enveloppe equation for the laser propagation, is fully
justified. In the present calculation we used an improved version
of WAKE, named WAKE\_EP \cite{Parakar_pop2013}, in which the fields
generated by the relativistic electrons bunch are taken into account
following the same procedure as in \cite{Morshed_Antonsen_2010},
where it was demonstrated that, for conditions similar to our cases,
WAKE-EP results are in good agreement with full PIC ones. In our conditions,
WAKE-EP is, at least, three orders of magnitude faster than a PIC
code.

\subsection{Beam loading effects}

\begin{figure}[htb]
\includegraphics[width=8cm]{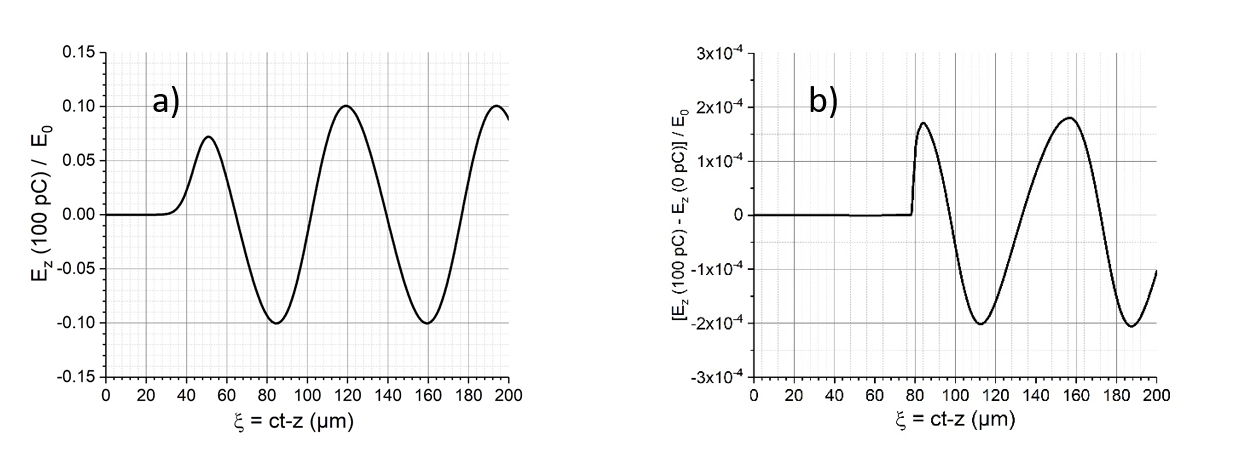}

\caption{Longitudinal electric field $E_{z}$ on axis in units of $E_{0}=m_{e}c\omega_{p}/e$
at 0.5 cm after the laser focal plane. In Fig. a the electron bunch
has a charge of 100 pC, in Fig. b, is reported the difference in $E_{z}$
with and without beam loading. \label{fig:figBeamLoading}}
\end{figure}

We have investigated whether the beam loading effect can have a significant
contribution in our conditions. Starting from an ASTRA file corresponding
to an optimal electron bunch configuration described in \cite{KeWangEAAC2017},
we have performed two calculations. One with no beam loading (total
charge of 0 pC) and a second one with a total charge of 100 pC. In
Fig. \ref{fig:figBeamLoading}-a we have reported the value of the
longitudinal field $E_{z}/E_{0}$ for the 100 pC case at a position
close to the focal plane of the laser, $E_{0}$ being given by $E_{0}=m_{e}c\omega_{p}/e$,
while $m_{e}$ and $-e$ are the mass and charge of an electron, $c$
is the speed of light in vacuum and $\omega_{p}$ is the plasma frequency;
in our conditions $E_{0}$ = 430 MeV/cm. In Fig. \ref{fig:figBeamLoading}-b
is reported, for the same position, the difference in the longitudinal
field values with and without beam loading. We observe that this difference,
which yields the field generated by the electron bunch, is non-zero
only at the position of the electron bunch and behind it, the electron
bunch generating its own plasma wave. The amplitude of this wave is
however quite small, its amplitude being 500 times smaller than the
one of the full plasma wave. At 10 pC it will be even 10 times smaller.

From these results, we can conclude that, in our conditions the beam
loading effect can be neglected. Note that in a WAKE-EP calculation,
when including the beam loading effect, the computing time increases
by nearly a factor of five, because one needs a high number of macro-particles
in order to reduce the numerical noise. Without beam loading, the
electrons trajectories can be calculated in a fixed external field,
which can be determined independently. This favorable feature has
been implemented in the numerical code WakeTraj in which the calculation
of the electrons trajectories is performed through a post-treatment
of one WAKE-EP calculation. Practically, during a WAKE-EP calculation,
tables of values of the field amplitude on a grid are periodically
save on disk. A WakeTraj calculation will use these tables to extrapolate
the field values at the electron positions. In the quasi-linear regime,
the typical length of variation of the field amplitude during propagation
is given by $Z_{R}$, which is rather large (1 cm in our case). We
have verified that with a propagation length of $Z_{R}/20$, between
two disk records, we get nearly no-difference between the WAKE-EP
and the WakeTraj results. In WakeTraj, there is also the possibility
to determine the electric field by applying linear theory formulas
\cite{esarey_physics_2009}, allowing to quantify the contribution
of non-linear effects, as it will be discussed in the next section.

\section{Results of parametric studies}
\label{sec:parametric}
\subsection{Optimization of the LPA}

\begin{figure}[htb]
\includegraphics[width=8cm]{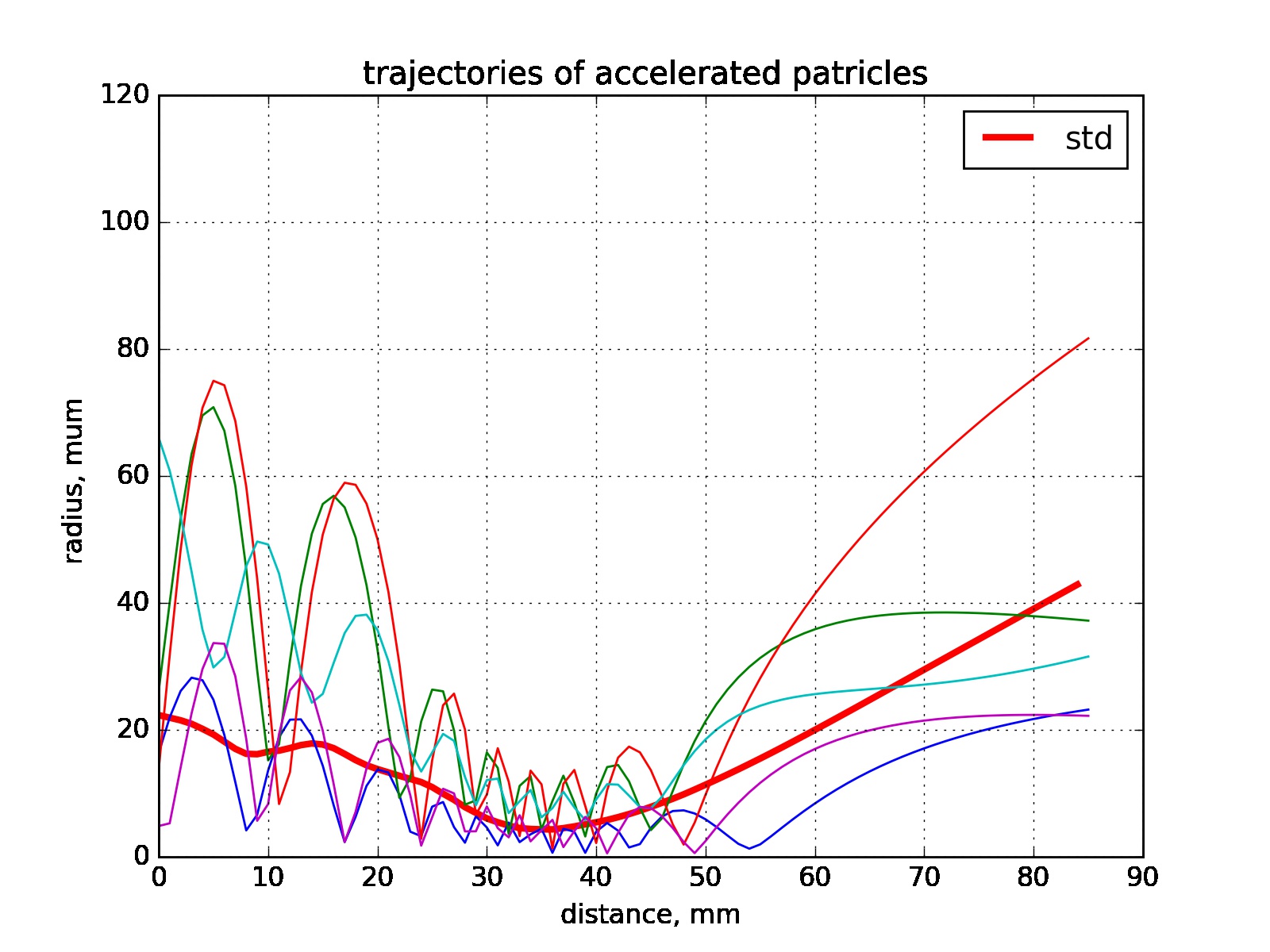}

\caption{Selected electron trajectories during the LPA process. The radius
is the distance to the laser axis. The thick red line represented
the average value of this radius.\label{fig:fewTrajectories}}
\end{figure}

Our three main criteria for optimizing the electron plasma interaction
process is to obtain the largest amount of trapped electrons, with
the highest energy and the minimum energy spread. This is accomplished
by starting the interaction process well before the laser focal plane,
where the longitudnal plasma field is relatively small, but the transverse
one is large enough to focus the electron bunch. This is illustrated
in Fig. \ref{fig:fewTrajectories}, in which few electrons trajectories
are represented, for a laser focal plane situated at 3 cm of the plasma
entrance. We observe first that even electrons with a relatively large
initial radius can be captured by the plasma wave. This is due to
the fact that at 3 cm from the focal plane, the laser waist reaches
160 \textmu m. We observe also that, during the focusing phase, the
electrons perform some betatron oscillations, showing that the focusing
is not too strong, so that there will be no large increase of the
electron emittance. Finally we can also see that at the end of the
interaction there is some re-focusing of the electrons due to the
transverse part of the plasma field, which yields a reduction of the
final divergence.

Although the electron bunch has been efficiently compressed before
entering the plasma (see \cite{KeWangEAAC2017}) its duration should
be reduced before being accelerated by the highest fields close to
the focal plane. Therefore, the focusing phase is also used to realize
a phase rotation in the plane $\xi,p_{z}$, with $\xi=ct-z$, $z$
being the position on the laser axis, $t$ the duration of propagation
and $p_{z}$ the electron momentum, which value is, in our relativistic
regime, close to the Lorentz factor $\gamma_{e}$.

\begin{figure}[htb]
\includegraphics[width=8cm]{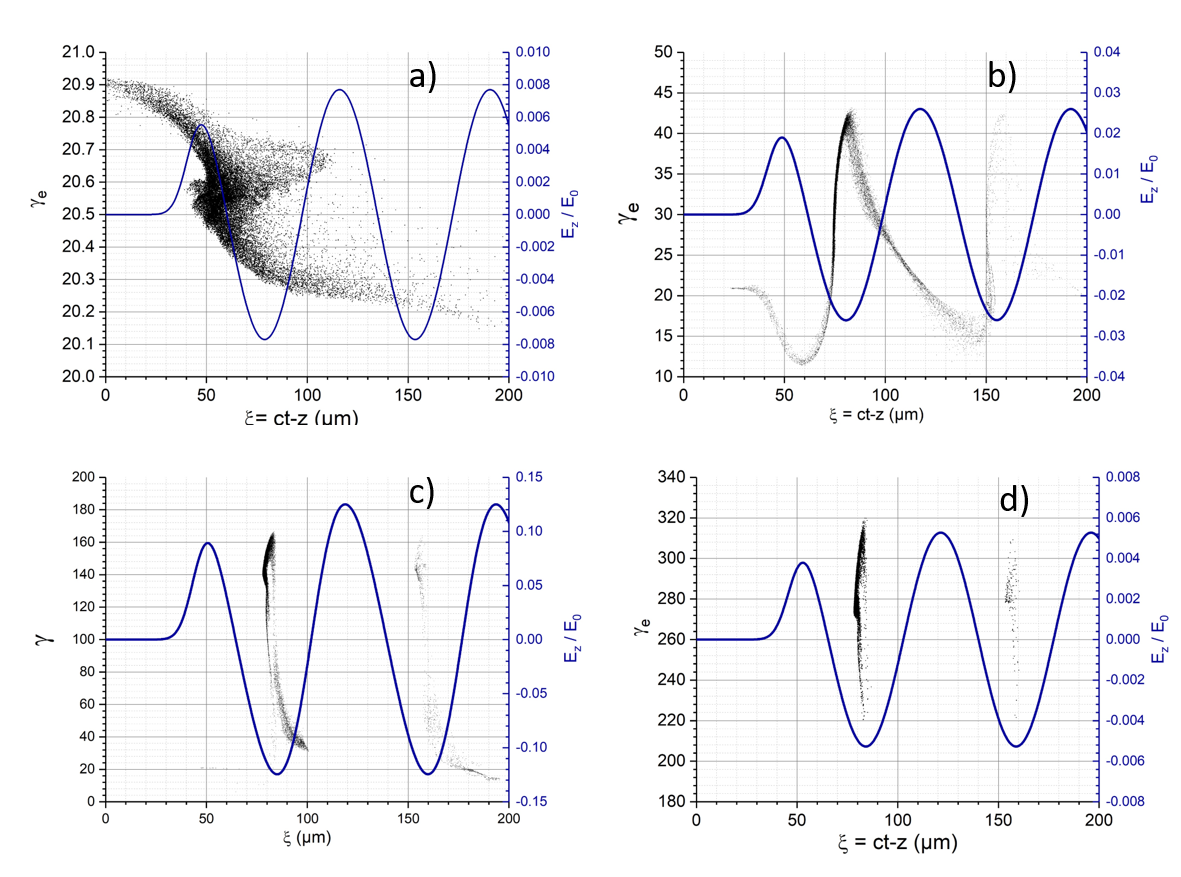}

\caption{Lorentz factor of the electron versus their longitudinal positions
(black points and left axis) and longitudinal electric field in reduced
units (blue curve, right axis) at four different distances of propagation.
entrance of the plasma (a); after 2 cm (b); at the focal plane (c)
and at the exit of the plasma (d). The focal plane is situated at
4 cm from the entrance of the plasma and the total cell length is
9 cm.\label{fig:PhaseSpace}}
\end{figure}

In Fig. \ref{fig:PhaseSpace}, is reported the phase space ($\xi,\gamma_{e})$
of the electron (left axis) together with the values of $E_{z}$ (right axis)
at four different positions. The front part of the laser is at $\xi\sim0$.
At the entrance of the plasma (Fig.\ref{fig:PhaseSpace}-a), and at
the central position of the bunch the longitudinal field is close
to 0, which corresponds to a maximum of the focusing transverse field.
Note that positive values of $E_{z}$ correspond to an accelerating
field, while the transverse field $E_{r}$ has a difference of phase
of $\pi/2$ with it. The first half period behind the laser is focusing,
and the second one defocusing. We observe that around its central part,
the electron bunch has a relatively long tail. A part of this tail,
interacting with a defocusing field will be lost, a second small
part will be trapped in the second period. After 2 cm of propagation
(Fig.\ref{fig:PhaseSpace}-b), some rotation of the phase space has
occurred, reducing the duration of the main part of the bunch. We see
that the bunch is situated close to the back of the first period,
where the accelerating field is maximum and its derivative is minimum.
At the focal plane, ( Fig.\ref{fig:PhaseSpace}-c)) here situated
4 cm from the plasma entrance, a short bunch of high energy electrons
is clearly identified. As this bunch is situated at a position where
the variation of $E_{z}$ is small, this bunch will be further accelerated
up to the plasma exit with a small increase of his energy spread,
as seen in ( Fig.\ref{fig:PhaseSpace}-d). 
Emittance of the injected electron bunch at the plasma entrance are $E_x= 1.3$ mm*mrad, $E_y = 1.4$ mm*mrad, the emittance at the exit of the plasma $E_x = 2.2$ mm*mrad, $E_y = 2.4$ mm*mrad for the horizontal and vertical planes respectively. One of the reasons of the emittance growth is related to multi-bunching, which takes place during the electrons capturing and acceleration by consecutive periods of wake field (See Fig.\ref{fig:PhaseSpace}-d).

As can been seen from the results of Fig.\ref{fig:PhaseSpace} the focusing
phase has a strong influence on the final properties of the bunch
electrons. For a given configuration two parameters can be used to
control the dynamics of the electrons in this phase: the laser focal
plane position and the delay between the incoming electron bunch and
the laser pulse. These two points are investigated in the next two
sections.

\subsection{Influence of the laser focal plane position}

\begin{figure}[htb]
\includegraphics[width=8cm]{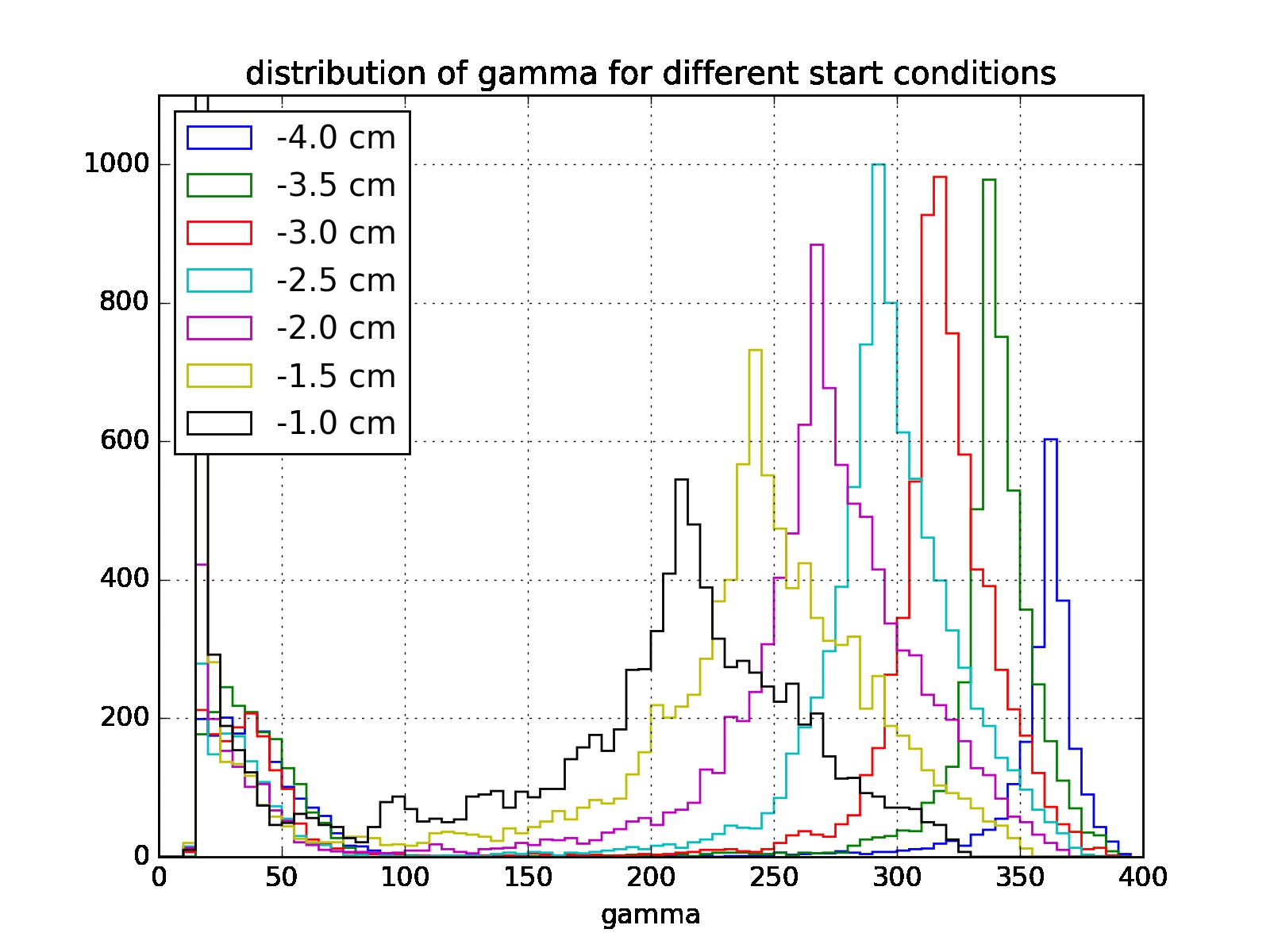}

\caption{Energy distribution of the bunch electrons in units of the Lorentz
factor, for different starting positions relative to the laser focal
plane.\label{fig:StartConditionHist}}
\end{figure}

In this sub-section we analyze the influence of the focal plane position.
Test electron bunch of about 20\% longer than the plasma wavelength is used in order to uniformly fill the plasma period with electrons and reduce the sensitivity of the results to the delay between the laser and the electron bunch, which is discussed in the next section 3.3.
In Fig. \ref{fig:StartConditionHist} are represented the energy distribution of the electrons at the exit of the plasma cell for a focal plane distance $d_F$ from the entrance of the plasma ranging from 1 cm up to 4 cm. The total plasma cell length is fixed to be 9 cm, the delay between the laser and the electron bunch is kept fixed.
Although the largest accelerating fields are obtained
close to the focal plane, we observe a significant modification of
the energy at the peak of the distribution even at distances of several
$Z_{R}.$ This can be explained by looking at the ratio between the
focusing $d_{F}$ and the dephasing $d_{\varphi}$ lengths. $d_{\varphi}$
is defined as the required transport distance so as to change the
relative distance between the laser and the electron by $\lambda_{p}/4,$ that
is to go from the position where the accelerating field is zero up
to the point where it is maximum. At 10 MeV, $d_{\varphi}=1.6$ cm.
Starting at large distances, $d_{F}\gg d_{\varphi}$, the electrons
have the time, before being strongly accelerated, to slip up to the
position of maximum accelerating field, at it was observed in Fig.
\ref{fig:PhaseSpace}(b-c). However, if this distance is too large,
a significant part of the electrons, in the back of the bunch, will
go into the defocusing zone and will be lost. In this case the final
energy distribution is rather sharp, but the accelerated charge is
reduced. This is the case of the blue curve in Fig. \ref{fig:StartConditionHist},
corresponding to a focal plane at 4 cm of the cell entrance. On contrary,
if we start close to the focal plane, $d_{F}<d\varphi,$ the electrons
are quickly accelerated before changing their relative positions.
As they are not initially situated at the maximum of the accelerating
field, the final acceleration will be lower. Furthermore, the phase
rotation mentioned in the previous sub-section cannot occur, yielding
a large energy spread, as observed in the black curve of \ref{fig:StartConditionHist}
corresponding to a 1 cm distance of the focal plane.

\begin{figure}[htb]
\includegraphics[width=8cm]{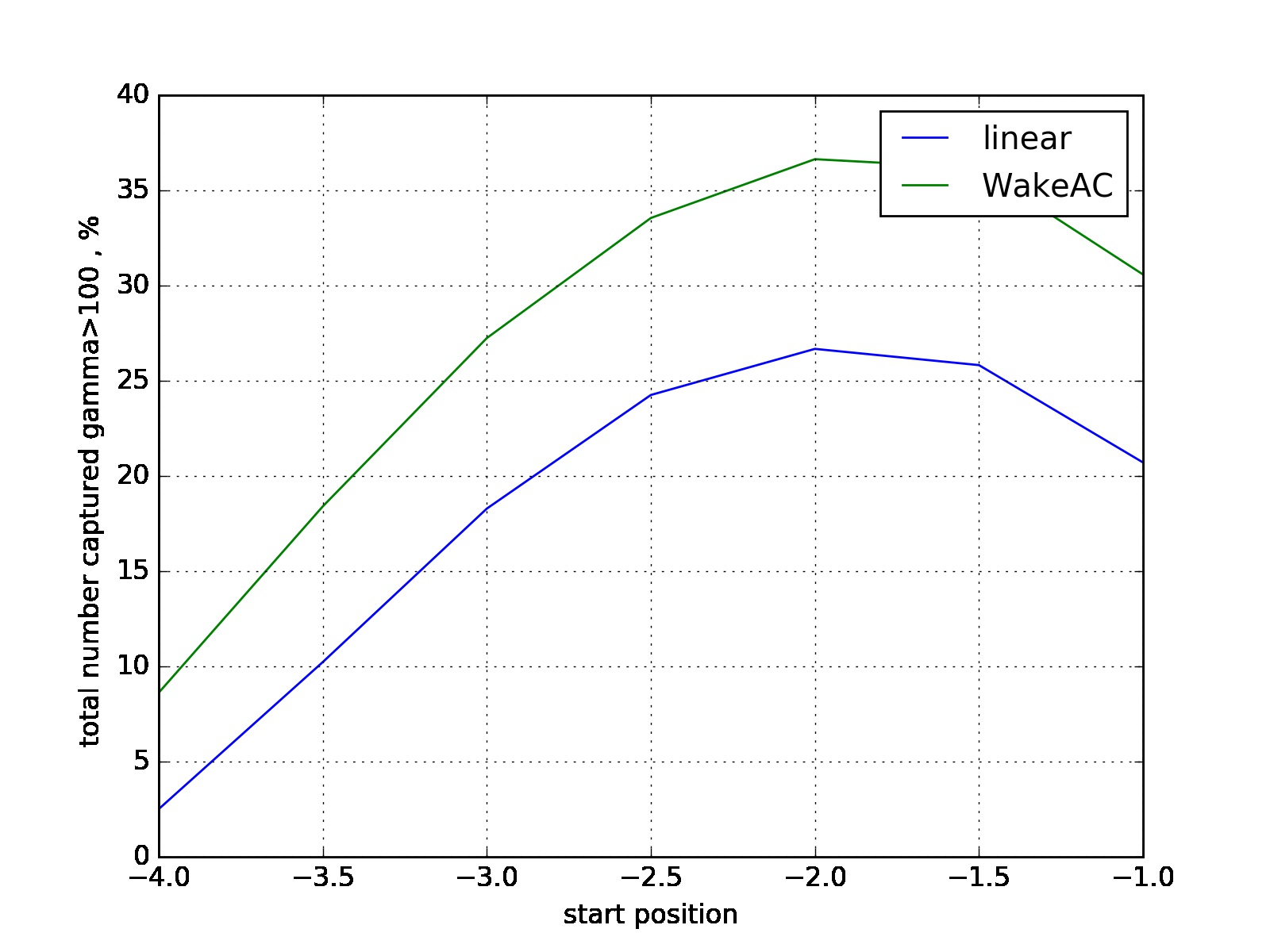}

\caption{Relative number of captured electrons, for different starting positions relative to the laser focal
plane.\label{fig:NumberCapturedElectrons}}
\end{figure}

In order to analyze the importance of the non-linear effects the same
calculations were also performed using the linear expression of the
electric field. In Fig. \ref{fig:NumberCapturedElectrons} are represented
the percentage of electron accelerated at $\gamma_{e}\geq100$, calculated
by the WakeTraj code using either the WAKE-EP electric field or the
linear theory one. We can observe a significant difference between
the two results. In fact, in our case, non-linear effects lead to
a slight increase of the dimension of the accelerating zone, which
increases the amount of trapped electrons and also their energies.
This increase is larger at large values of $d_{F},$ when the percentage
of captured electrons is the smallest. In Fig. \ref{fig:NumberCapturedElectrons},
we observe also that the amount of accelerated charge is decreasing
when the plasma entrance becomes close to the focal plane. This is
due to the fact that, close to the focal plane, the waist of the laser
has its lowest value, as a consequence, electrons with the largest
radius can no more be trapped.

\subsection{Influence of the delay between the laser and the electron bunch}

In Fig. \ref{fig:StartConditionHist} we can observe that the energy
spread is decreasing when $d_{F}$ is increased. Further optimization
requires to increase also the amount of accelerated electrons. A first
step has been performed through an additional compression of the electron
bunch, as obtained in the fully optimized configuration of the ESCULAP
beam line described in \cite{KeWangEAAC2017}. The second step relies
on a careful optimization of the delay between the maximum amplitude
of the laser and the maximum of the electron bunch density. This delay
has been modified between 14 and 93 fs. A summary of the obtained
results is reported in Table \ref{tab:tab1} for an injected charge
of 10 pC. From this table we see that the reduction of the bunch duration
together with an optimization of the delay provide a very strong increase
of the accelerated charge reaching up to 86 \% of the injected one,
with an rms energy spread as low as 4.1 \%, a duration that has be
reduced by a factor 10 and a divergence around two mrad.

\begin{table}[tbh]
\centering \caption{Main properties of the electron bunch in terms of the delay between
the laser and the bunch: average energy $<E>$, dispersion in energy
$\sigma_{rms}(E)/E$, divergence $\theta_{rms}$ and duration $\tau_{FWHM}$}
\label{tab:tab1} %
\begin{tabular}{|c|c|c|c|c|c|}
\hline
delay & charge & <$E$> & $\sigma_{rms}(E)/E$ & $\theta_{rms}$ & $\tau_{rms}$\tabularnewline
\hline
\hline
fs & pC & MeV & \% & mrad & fs\tabularnewline
\hline
13.85 & 8.6 & 139.5 & 5 & 2.4 & 5.2\tabularnewline
\hline
23.76 & 8.4 & 140.7 & 4.5 & 2.2 & 6.0\tabularnewline
\hline
33.67 & 8.3 & 141.8 & 4.2 & 1.9 & 6.7\tabularnewline
\hline
43.57 & 8.2 & 142.6 & 4.1 & 1.8 & 7.5\tabularnewline
\hline
53.48 & 7.9 & 143 & 4.2 & 1.7 & 7.5\tabularnewline
\hline
63.39 & 7.6 & 143.7 & 4.4 & 1.6 & 7.1\tabularnewline
\hline
73.30 & 7.1 & 144.6 & 4.6 & 1.6 & 6.9\tabularnewline
\hline
83.21 & 6.6 & 146.1 & 4.9 & 1.7 & 7.8\tabularnewline
\hline
93.12 & 5.9 & 147.4 & 5.0 & 1.7 & 8.5\tabularnewline
\hline
\end{tabular}
\end{table}

\begin{figure}[htb]
\includegraphics [width=8cm] {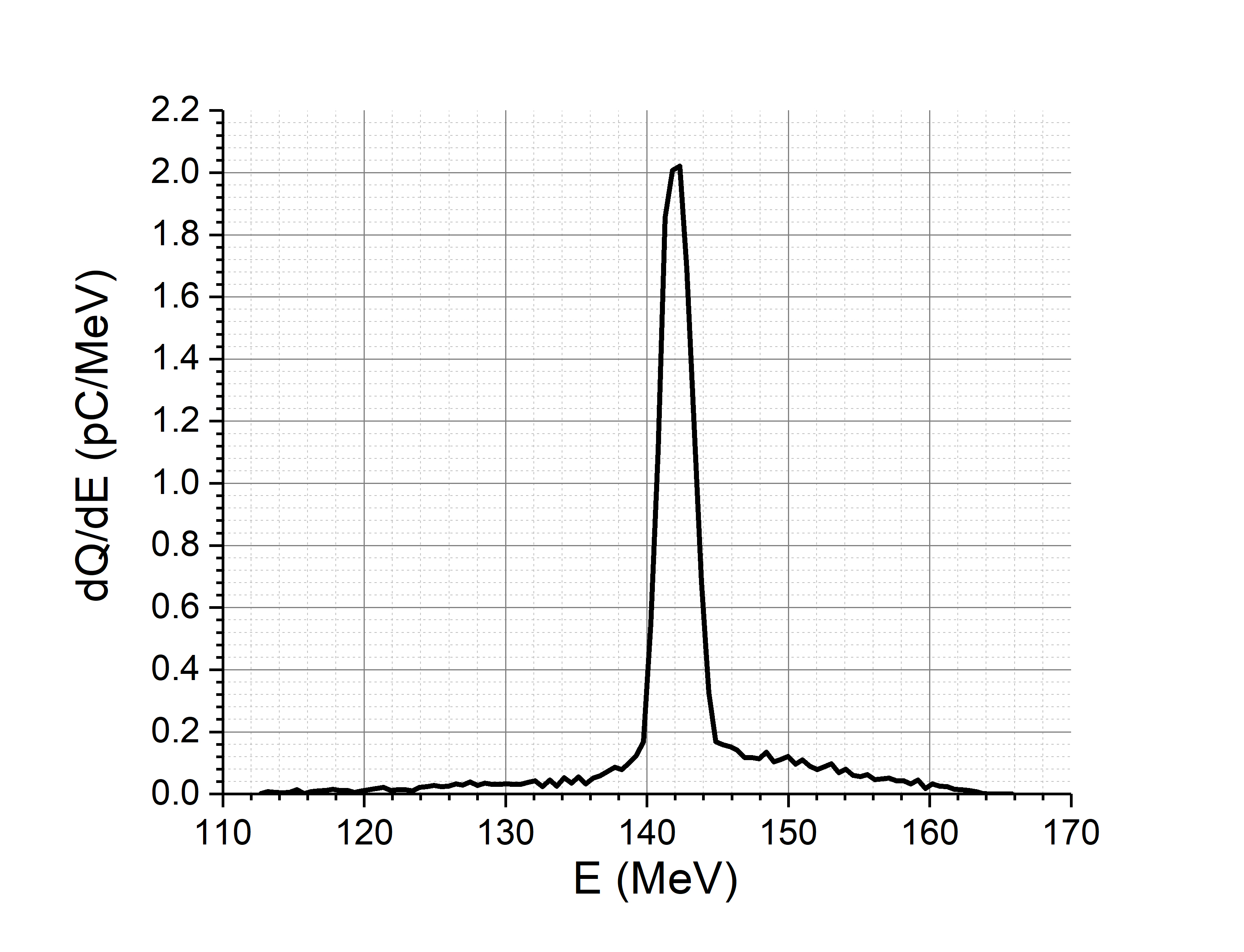}

\caption{Energy spectrum of the accelrated electrons for a laser-bunch delay
of 43.57 fs. .\label{fig:BestEnergyDistrib}}
\end{figure}

In Fig. \ref{fig:BestEnergyDistrib} is
reported the energy spectrum of the accelerated bunch for a delay
of 43.57 fs. We observe a very sharp peak, with a maximum value of
2 pC/MeV and a relative FWHM width of only 2 \%.

\section{Conclusion}
A parametric study of the laser plasma acceleration process have been
performed within the framework of the ESCULAP project. This has been
done using the numerical code WakeTraj, which allows through very
fast numerical calculations to investigate the influence of several
parameters. Results concerning the laser focal plane position and
the delay between the laser and the electrons bunch have presented.
From these results an optimized configuration has been found which
leads to an efficient coupling between the 10 pC, 10 MeV electron
bunch and the plasma wave generated by the 2 J LASERIX laser. Using
a 9 cm long gas cell more than 75 \% of the bunch electrons can be
trapped in the plasma wave and accelerated up to an average energy
close to 150 MeV, with a dispersion in energy as small as 4 \% a divergence
of 1.7 mrad and a duration of 7.5 fs, which offers interesting perspectives,
in particular for further acceleration in a guiding structure.

\bibliographystyle{elsarticle-num}
\bibliography{ALP}

\begin{thebibliography}{1}
\expandafter\ifx\csname url\endcsname\relax
  \def\url#1{\texttt{#1}}\fi
\expandafter\ifx\csname urlprefix\endcsname\relax\def\urlprefix{URL }\fi
\expandafter\ifx\csname href\endcsname\relax
  \def\href#1#2{#2} \def\path#1{#1}\fi

\bibitem{NiDelerueEAAC2017}
E.~Baynard, C.~Bruni, K.~Cassou, V.~Chaumat, N.~Delerue, J.~Demailly,
  D.~douillet, N.~El~Kamchi, D.~Garzella, O.~Guilbaud, S.~Jenzer, S.~Kazamias,
  V.~Kubytskyi, P.~Lepercq, G.~Lucas, O.~Neveu, M.~Pittman, R.~Prazeres,
  H.~Purvar, D.~Ros, K.~Wang, The esculap project at orsay: External injection
  of low energy electrons in a plasma., in: these proceedings, 2017.

\bibitem{KeWangEAAC2017}
K.~Wang, E.~Baynard, C.~Bruni, K.~Cassou, V.~Chaumat, N.~Delerue, J.~Demailly,
  D.~douillet, N.~El~Kamchi, D.~Garzella, O.~Guilbaud, S.~Jenzer, S.~Kazamias,
  V.~Kubytskyi, P.~Lepercq, G.~Lucas, O.~Neveu, M.~Pittman, R.~Prazeres,
  H.~Purvar, D.~Ros, Longitudinal compression and transverse matching of
  electron bunch for external injection lpwa at esculap, in: these proceedings,
  2017.

\bibitem{esarey_physics_2009}
E.~Esarey, C.~B. Schroeder, W.~P. Leemans,
  \href{http://link.aps.org/doi/10.1103/RevModPhys.81.1229}{Physics of
  laser-driven plasma-based electron accelerators}, Reviews of Modern Physics
  81~(3) (2009) 1229--1285.
\newblock \href {http://dx.doi.org/10.1103/RevModPhys.81.1229}
  {\path{doi:10.1103/RevModPhys.81.1229}}.
\newline\urlprefix\url{http://link.aps.org/doi/10.1103/RevModPhys.81.1229}

\bibitem{vay_recent_2016}
J.~L. Vay, R.~Lehe, H.~Vincenti, B.~B. Godfrey, I.~Haber, P.~Lee,
  \href{http://www.sciencedirect.com/science/article/pii/S0168900215016046}{Recent
  advances in high-performance modeling of plasma-based acceleration using the
  full {PIC} method}, Nuclear Instruments and Methods in Physics Research
  Section A: Accelerators, Spectrometers, Detectors and Associated Equipment
  829 (2016) 353--357.
\newblock \href {http://dx.doi.org/10.1016/j.nima.2015.12.033}
  {\path{doi:10.1016/j.nima.2015.12.033}}.
\newline\urlprefix\url{http://www.sciencedirect.com/science/article/pii/S0168900215016046}

\bibitem{mora_kinetic_1997}
P.~Mora, J.~Thomas M.~Antonsen,
  \href{http://dx.doi.org/10.1063/1.872134}{Kinetic modeling of intense, short
  laser pulses propagating in tenuous plasmas}, Physics of Plasmas 4~(1) (1997)
  217.
\newblock \href {http://dx.doi.org/10.1063/1.872134}
  {\path{doi:10.1063/1.872134}}.
\newline\urlprefix\url{http://dx.doi.org/10.1063/1.872134}

\bibitem{Parakar_pop2013}
B.~S. Paradkar, B.~Cros, P.~Mora, G.~Maynard, Numerical modeling of multi-gev
  laser wakefield electron acceleration inside a dielectric capillary tube,
  Physics of Plasmas 20 (2013) 083120.

\bibitem{Morshed_Antonsen_2010}
S.~Morshed, T.~M. Antonsen, J.~P. Palastro, Efficient simulation of electron
  trapping in laser and plasma wakefield acceleration, Physics of Plasmas 17
  (2010) 063106.

\end{thebibliography}

\end{document}